\documentclass[aps,pre,preprint,groupedaddress]{revtex4}

\begin{document}


\title{Thermodynamics of non-equilibrium steady states}

\author{Glenn C. Paquette}
\email[]{paquette@scphys.kyoto-u.ac.jp}
\affiliation{Department of Physics, Graduate School of Science, Kyoto University, Kyoto 606-8502, Japan }

\date{\today}

\begin{abstract}
We consider the problem of constructing a thermodynamic theory of non-equilibrium steady states as a formal extension of the
equilibrium theory. Specifically, studying a particular system, we attempt to construct a phenomenological theory describing the
interplay between heat and mechanical work that takes place during operations through which the system undergoes transitions
between non-equilibrium steady states. We find that, in contrast to the case of the equilibrium theory, apparently, there exists no
systematic way within a phenomenological formulation to describe the work done by the system during such operations. With this
observation, we conclude that the attempt to construct a thermodynamic theory of non-equilibrium steady states in analogy to the
equilibrium theory has limited prospects for success and that the pursuit of such a theory should be directed elsewhere. 

\end{abstract}

\maketitle

\section{Introduction}
{\label{intro}}

In this paper, we examine the proposition of constructing a thermodynamic theory of non-equilibrium steady states
in a form that mimics that of the equilibrium theory. There are a number of existing works that formulate various extensions
of the equilibrium theory to non-equilibrium systems, and some of these formulations are applicable to non-equilibrium steady states,
in particular those in \cite{jou, eu, keizer, oono1, oono2, komatsu}. However, none of these works yields results of general validity in the case of
non-equilibrium steady states, as each is limited to either systems close to equilibrium or systems in which the asymptotic
states differ from equilibrium only trivially \cite{fn1}. Here, we investigate the
prospect for constructing such an approach that is of general applicability.

The theoretical framework of equilibrium thermodynamics is fundamentally
a formalism for describing the manner in which thermal and mechanical forms of energy are converted into one another in processes that
connect equilibrium states. In this paper, we consider the construction of such a framework for non-equilibrium steady states.
More precisely, we consider the formulation of a kind of thermodynamic potential
in terms of which we can understand the interplay between heat and mechanical work
in operations connecting non-equilibrium steady states, and we attempt to obtain a generalized second law valid for such systems. 
Through this investigation, we reach the conclusion that a theory of this type is of quite limited use.

As we show here, the main difficulty faced in the attempt to construct a thermodynamic theory of non-equilibrium steady states similar to the equilibrium
theory is that when we consider operations in which
a system undergoes a transition from one steady state to another through the variation of the externally adjustable mechanical parameters (work coordinates),
the work done by the system cannot be easily characterized. There are two particular problems that
we encounter in this regard. First, in general, for a quasi-static, purely isothermal operation,
the work done by the system through the work coordinates as it goes  from one specific steady state to another is not independent of the path taken
among these coordinates. Second, for a given path, the amount of work
done by the system through the work coordinates is not necessarily maximal in the case of a quasi-static operation, and there is no apparent way
to determine when a quasi-static operation maximizes
this work and when it does not. From these results, obtained for a particularly simple system, it is surmised that
the formulation of a generally useful thermodynamic theory of non-equilibrium steady states that describes the conversion of heat
into mechanical forms of energy is an extremely difficult problem at best. We thus believe that the attempt to construct
a phenomenological theory of non-equilibrium steady states should be directed elsewhere.
This conclusion is consistent with the observation that fundamental
advances in physics result not from the formal extrapolation of existing theories but from the discovery of simple, novel relations involving readily measured quantities.

\section{Preliminaries}
{\label{prelim}}

In this section we introduce the system studied and present the expressions for a number of quantities
that are used repeatedly throughout the paper.

\subsection{The system}

In order to demonstrate the points of this paper in the clearest manner, we consider a very simple system, consisting of a single Brownian
particle in one spatial dimension. Throughout this work, we regard this particle as the `system' and
the fluid medium in which it is suspended as a heat bath \cite{fn1.1}.
We study the case in which the dynamics of the system can be described by a Langevin equation of the form
\begin{equation}
\gamma \dot{x}(t) = {\cal F}(x(t);\alpha(t)) + \xi(t) \; ,
\label{langevin}
\end{equation}
where $x$ denotes the position of the particle, $\gamma$ is a friction constant, $\alpha(t)$ is an externally variable parameter on which the systematic force ${\cal F}$ depends,
and $\xi(t)$ represents a Gaussian noise of intensity $2 \gamma T$,
with $T$ the temperature of the heat bath
(choosing units in which the Boltzmann constant is unity). 

Now, we wish to consider the situation in which the deterministic force, ${\cal F}(x; \alpha)$, takes the form
${\cal F}(x; \alpha) = F(x; \alpha) + f$, where $F(x; \alpha)$ is a conservative force derived from some potential
$V(x;\alpha)$, and $f$ is a non-conservative force that is uniform in space.
In general, if $\alpha$ is held fixed,
a system of this kind will converge to a non-equilibrium steady state characterized by this
value of $\alpha$ in the large time limit.

Applying the stochastic Liouville equation to (\ref{langevin}), it is found that the dynamics of
the particle number probability distribution function, $\rho(x,t)$, are described by the following
Smoluchowski equation \cite{fn2}:
\begin{equation}
\gamma \dot{\rho}(x,t) = \partial / \partial x  \left[ T \partial / \partial x
-  {\cal F}(x; \alpha(t)) \right] \rho(x,t) \; .
\label{equationofmotion}
\end{equation}
Note that the quantity $x$ in this equation and that in (\ref{langevin}) have different meanings, as here it is an
independent variable, whose domain is the configuration space.
We consider the case in which the system is subject to the periodic boundary conditions
$\rho(L,t) = \rho(0,t)$.

\subsection{Important quantities}

In this paper, we investigate processes in which $\alpha$ is varied in time. To make the following treatments as clear as possible,
we now present the important quantities used throughout the paper.

First, at any given position and time, there will be some current in the system, $j(x,t)$, representing the
rate of particle flow through the point $x$ at time $t$ in the direction of increasing $x$ \cite{fn2.1}. In the overdamped case considered
here, this quantity is given by
\begin{equation}
j(x,t) = \left[\gamma^{-1} {\cal F} - D \frac{\partial}{\partial x} \right] \rho(x,t) \; , \label{current}
\end{equation}
where $D \equiv T/\gamma$ is the diffusion coefficient.
We write the spatial average of $j(x,t)$ as $\overline{ j }(t)$. (Throughout the paper, we use
the overline to represent the spatial average.) From (\ref{current}), we immediately obtain
\begin{equation}
\overline{ j } = \gamma^{-1} c_0 f + \gamma^{-1} \overline{ F \rho } \; , \label{averagecurrent}
\end{equation}
where $c_0 \equiv L^{-1}$.
Note that in a steady state, we have $j = \overline{ j }$. We represent the current in this special case by $j^{\rm{ss}}$.
If the system is not in a steady state, there will generally exist some non-uniform current, $j^{\rm{non}}  \equiv j - \overline{ j }$.
  
At any time, the energy of the system, $U$, is given by $U(t) = \int dx V(x;\alpha(t)) \rho(x,t)$.
(Throughout the paper, all integrals over the spatial coordinate are done over the entire
system, and thus, for simplicity, we omit the limits of integration.)
Next, note that the rate of work done by the external agent when changing $\alpha$, $w_F(t)$, is
given by $\dot{\alpha}\frac{\partial}{\partial \alpha} \int dx V(x ; \alpha) \rho(x,t)$.
%
%
Then, because $\rho(x,t)$ is not an explicit function of $\alpha$ and the only time dependence
in $V$ is that in $\alpha$, we have
\begin{equation}
w_F = \int dx \frac{dV(x;\alpha(t))}{dt}  \rho(x,t)\; . \label{wF}
\end{equation}
Next, consider the quantity $\int dx V(x; \alpha) \dot{\rho}(x,t)$. Using the conservation
of particle number and some simple manipulations, it is readily shown
that this is equal to $- \int dx F(x;\alpha) j(x,t)$. Then, noting the form of $F(x;\alpha)$, obviously,
it is also equal to $- \int dx F(x;\alpha) j^{\rm{non}}(x,t)$. This quantity represents the opposite of the total rate of work
done by means of the force $F$ on the Brownian particle. Hence, because we are considering the case in which the motion of
the particle is overdamped, this must be equal to the rate of work done by the heat bath on the non-uniform component
of the current. Following Sekimoto \cite{sekimoto}, we interpret this
as the rate of transfer of a type of heat, which we write $q^{\rm{non}}$. Thus, we have
\begin{equation}
q^{\rm{non}} \equiv - \int dx F(x;\alpha) j(x,t) \; . \label{qnon}
\end{equation}
Combining the above results, we obtain the following:
\begin{equation}
\frac{d U}{dt} = w_F + q^{\rm{non}} \; . \label{changeofenergy}
\end{equation}

The rate of work done by means of the non-conservative force, $w_f$, is given by
\begin{equation}
w_f = L f \overline{ j } \; . \label{wf}
\end{equation}
Once again, because the motion of the particles is overdamped, we have 
\begin{equation}
q^{\rm{uni}} = - w_f \; , \label{quni}
\end{equation}
where $q^{\rm{uni}}$ is the rate of transfer of ``uniform heat," which is defined as the rate of
work done by the heat bath on the uniform component of the current. Because $q^{\rm{uni}}$ and $w_f$ always cancel,
they play no direct role in the following analysis.

We express the time integrals of the quantities $w_F$, $w_f$, $q^{\rm{non}}$ and $q^{\rm{uni}}$ over an entire
process by $W_F$, $W_f$, $Q^{\rm{non}}$ and $Q^{\rm{uni}}$.

\section{Direct generalization of equilibrium formalism}
\label{eqform}
\subsection{Motivation of approach}

The formalism of equilibrium thermodynamics is based on a thermodynamic potential that is  a state function. Owing to the special properties
of equilibrium systems, as embodied in this thermodynamic potential, the performance of work can be described in a quasi-mechanical
manner, because the work done by a system in an infinitesimally slow operation connecting any two initial and final states is simply given by the
difference between the values of the thermodynamic potential at these endpoints. Hence, in this case, the conversion of heat into mechanical forms
of energy can also be described within this quasi-mechanical framework, entirely in terms of thermodynamic state variables. The usefulness of the formalism based on the thermodynamic potential
follows from these properties and the fact that, with regard to the conversion of heat into mechanical forms of energy, there is a simple relation between the
behavior of the system in quasi-static and non-quasi-static operations, with the transfer of heat to the system and the performance of work by the system always being maximized by the former.

Noting that non-equilibrium steady states are realized as asymptotic fixed-point solutions in the time evolution
of a large class of systems, it is natural to conjecture that such states, like equilibrium states, correspond to minima
of some quantity characterizing the state space, i.e., some kind of non-equilibrium thermodynamic potential.
It is thus reasonable to seek a formalism of non-equilibrium steady states analogous to that of
equilibrium states, in which a description of the performance of work and the exchange of heat is
expressed in terms of state variables.
In fact, there exist attempts to construct such a formalism \cite{keizer, oono1, oono2, komatsu}, but as shown in this section, obtaining
a general theoretical framework for non-equilibrium steady states based on a thermodynamic potential
is perhaps unfeasible.

In the case of non-equilibrium steady states, in contrast to that of equilibrium states,
even for a quasi-static operation, the total transfer of heat cannot be accounted for within a mechanical
picture, as the steady state current results in the continual
dissipation of work into heat. Recognizing this, Oono and Paniconi \cite{oono1, oono2} proposed an approach to constructing
a theory of non-equilibrium steady state systems that begins by subtracting out this steadily dissipated
heat (what they call the ``housekeeping heat"). They conjectured that if this is done in the proper manner,
the remaining heat (the ``extra heat") will play a role like
that of the total heat in the equilibrium case. If these roles were perfectly analogous, it would
be possible to define a thermodynamic potential characterizing the quasi-mechanical behavior of the
system that remains when the housekeeping heat is ignored
in the case of an infinitesimally slow operation, and the deviation from this behavior in the
case of a finitely fast operation would be easily characterized.
In this section, we find that for some types of systems,
the idea of distinguishing between different types of heat does indeed facilitate
the construction of a thermodynamic potential.
It is thus seen that the intuitively appealing approach proposed by Oono and Paniconi
is useful. However, as shown in Appendix \ref{appqex}, for (\ref{equationofmotion}),
the component of the heat in the non-equilibrium case that is analogous to the total heat in the
equilibrium case is not the extra heat that they define but the non-uniform heat, $Q^{\rm{non}}$,
defined in the previous section. In general, these quantities are distinct for the system studied
here, due to the presence of a spatially uniform component of the extra heat (see Appendix \ref{appqex}).
Next, it is shown that, unlike in the equilibrium case, the deviation from quasi-mechanical behavior of the system
that appears for a finitely fast operation is not easily characterized. Finally, it is shown that, again unlike in the equilibrium case,
despite the fact that infinitesimally slow
operations are reversible (if we subtract the uniform heat), in general they are not path independent. For this reason, although it is
possible to define a thermodynamic potential given a particular path in parameter space, the
potentials corresponding to different paths will differ, and hence the thermodynamic potential is not a state function.
We thus conclude that the usefulness of a formalism based on a thermodynamic potential -- without the introduction
of a drastically different manner of thinking -- is doubtful.

\subsection{Perturbative solution to equation of motion}
\label{pert}

In the following, we consider an ensemble of experiments described by (\ref{equationofmotion})
that are carried out under the following conditions. First, the system is prepared by keeping
$\alpha$ fixed at some initial value $\alpha_{\rm{i}}$ over the time interval
$t \in (-\infty, t_{\rm{i}})$, for some finite value $t_{\rm{i}}$.
Then, $\alpha$ is varied in some manner over the interval $t \in [t_{\rm{i}} ,t_{\rm{f}}]$. (We assume that
$\alpha(t)$ is analytic for all $t \in (t_{\rm{i}}, t_{\rm{f}})$.) Finally,
$\alpha$ is held fixed again to some value $\alpha_{\rm{f}}$ for all $t > t_{\rm{f}}$, and the system is
allowed to relax to some new steady state. Thus, the operation carried out by changing
$\alpha$ induces a transition between the steady states corresponding to $\alpha_{\rm{i}}$ and $\alpha_{\rm{f}}$. We would like
to derive a solution $\rho(x,t)$ that describes the system for all $ t \in [t_{\rm{i}}, t_{\rm{f}}]$,
but because the nature of the experiment implies that there will be some nonanalytic behavior
at $t = t_{\rm{i}}$,
the solution for times close to
$t_{\rm{i}}$ will depend in a complicated
manner on these initial conditions. For this reason, obtaining a solution that is valid for this
entire time interval is not possible. However, because $\rho(x,t)$ represents an ensemble of random walks,
it is expected that as the time increases, all dependence on the initial conditions (i.e., the nonanalytic
behavior) will be lost \cite{fn3}.
Thus, we conjecture that there exists a unique form to which $\rho(x,t)$ will converge in the limit of large $t -t_{\rm{i}}$.
We expect that this solution will be simple, in the sense that the complicated dependence on the initial, non-analytic
behavior will have dissipated away, and that it will be universal, in the sense that it is the form to which $\rho(x,t)$
converges for all $\rho(x,t_{\rm{i}})$. In the parlance of renormalization group (RG) theory, this form represents an RG fixed point,
and it is often referred to as the `scaling form' solution, describing the `scaling' or `intermediate asymptotic' regime.
In the context of center manifold theory, this universal form represents the center manifold itself.
In the following, we derive a solution of this type.
Of course, in order for such a solution to be realized in an experiment of the type considered here, it is necessary that
$t_{\rm{f}}$ be sufficiently large.  We assume this to be the case. Actually, in the following, we assume that this scaling
regime solution represents
the behavior of the system over the entire interval $[t_{\rm{i}},t_{\rm{f}}]$, ignoring the transient behavior entirely.
While this introduces an error into several of the results
derived here (for example, in the total work done through $F$ over this interval), if we consider only
very long operations, the contribution of the transient, and hence the error in our results, should be negligible.
In fact, we estimate the error introduced by ignoring the transient when computing quantities such as the total work and
the total heat in Appendix \ref{apptrans}. Then, we use that result
in Section \ref{pert} and Appendix \ref{appqex} to show that the contribution of the transient
to each of the quantities computed there is negligible.

Throughout this paper, the
concept of the timescale of the variation of $\alpha$ plays an important role. For this reason, it
is convenient to introduce a dimensionless parameter, $\mu$, whose inverse characterizes this timescale, and we
write $\alpha$ as $\alpha(\mu t)$.
We also assume that the solution
to (\ref{equationofmotion}) can be written as an expansion in powers of $\mu$ in the form $\rho(x,t) =
\rho_0(x,t) + \mu \rho_1(x,t) + \mu^2 \rho_2(x,t) + \cdots $.
Here,
the quantities $\rho_n(x,t)$ are ${\cal O}(\mu^0)$
functions. In general, such a solution can be obtained by substituting this form into the
equation of motion and solving the resulting equations order by order in $\mu$, beginning with the ${\cal O}(\mu^0)$
equation.
Certainly, it is not always the case that the expansion obtained in
this manner converges, and, in particular, it is generally not possible to write such a solution for the
transient regime.
However, for sufficiently small $\mu$, this procedure should yield a valid solution in the intermediate asymptotic regime \cite{fn4}.

In Appendix \ref{apppert}, we explicitly construct a solution to (\ref{equationofmotion}). In this calculation, both
$\mu$ and $\alpha$ are regarded as small quantities, and we seek a solution to (\ref{equationofmotion})
perturbatively in both. (For convenience, we write $\alpha(\mu t) = b a(\mu t)$ and regard $b$ as a small quantity.
The calculation is then carried out as a formal perturbative expansion in $\mu$ and $b$.)
With the various conditions described above, and employing the perturbative method demonstrated in Appendix \ref{apppert},
such a solution can be constructed straightforwardly. However, the computation rapidly becomes complicated
as the orders in $\mu$ and $b$ increase, and for this reason, we carry out this perturbative calculation
only to first order in both of these quantities. Also, for simplicity, we consider the case in which the potential
takes the form $V(x;\alpha) = \alpha \cos(kx)$.

With the above assumptions, from the computation given in Appendix \ref{apppert}, to ${\cal O}(\mu^0,b^1)$, we obtain the following:
\begin{equation}
\rho^{(0)}(x,t) = \left[ 1 + b a (\mu t) \left( \sigma_+^0 \cos(kx) + \sigma_-^0 \sin (kx) \right) \right] c_0 \; .
\label{order0}
\end{equation}
Here we have $\sigma_+^0 = - k^2 T / \theta$ and $\sigma_-^0 = - k f/\theta$, with $\theta = k^2 T^2 + f^2$.
As noted in the previous section, $c_0$ is the average particle density. We also note here that throughout the paper, we
write as $\rho^{(n)} (x,t)$ an approximate form of $\rho(x,t)$ valid to ${\cal O}(\mu^n)$.

Before proceeding to the next order in $\mu$, we point out that $\rho^{(0)}(x,t)$ depends on
time only through $\alpha(\mu t)$ and that, in fact, we have the identity
$\rho^{(0)}(x, t) = \rho_{\rm{ss}}(x ; \alpha(\mu t))$, where $\rho_{\rm{ss}}(x ; \alpha(\mu t))$
is the steady state solution corresponding to $\alpha = \alpha(\mu t)$. This is important in the
analysis carried out below.

Then, as shown in Appendix \ref{apppert}, at ${\cal O}(\mu^1)$, we have the contribution
\begin{equation}
\rho_1(x,t) = b a^\prime(\mu t) \gamma 
\left( \sigma_+^1 \cos(kx) + \sigma_-^1 \sin (kx) \right) c_0
 \; ,
\label{order1}
\end{equation}
where $a^\prime \equiv da/d(\mu t)$, and we have $\sigma_+^1 = (k^2 T^2 - f^2)/\theta^2$ and $\sigma_-^1 = 2kT f / \theta^2$.

Combining the above results, we obtain the following form for $\rho(x,t)$, valid to ${\cal O}(\mu^1, b^1)$:
\begin{equation}
\rho^{(1)}(x,t) = c_0 \left\{ 1 + \left[ \sigma^0_+ \alpha + \gamma \sigma^1_+ \dot{\alpha} \right]\cos(kx) +
\left[ \sigma^0_- \alpha + \gamma \sigma^1_- \dot{\alpha} \right] \sin (kx) \right\} \; . \label{order1form}
\end{equation}

\subsection{Transitions between steady states: work, thermodynamic potential and entropy}

\subsubsection{Overview}

Here, we briefly describe a formalism analogous to that employed in the study of equilibrium systems that
can be developed for the system studied here. This formalism is then used to study the behavior of
various thermodynamic quantities in the type of experiments described above.

In the following, we show that
the steady state solutions of (\ref{equationofmotion})
are characterized by a thermodynamic potential, $\Phi$, from which the total work done by means of $F$
in the case of an infinitesimally slow
operation can be derived. We also show that the quantity playing the role of the entropy in this potential
is {\it not} the equilibrium entropy. Then, we demonstrate that the change of this generalized entropy
in an infinitesimally slow operation is equal to a particular type of heat absorbed by the system (namely, $Q^{\rm{non}}$) divided by the temperature.
Next, we find that
for a finitely fast operation, depending on the values of $f$, $T$ and $\alpha$, the inequality in the second-law-like
relation characterizing this system can be reversed. It is thus seen that the deviation from quasi-mechanical behavior of the system in
a finitely fast operation causing a transition between non-equilibrium steady states
cannot be characterized in a simple manner, unlike in the case of transitions between
equilibrium states. Finally, we show that if the adjustable parameter possesses multiple degrees of freedom,
even in the case of infinitesimally slow operations (and with an entirely isothermal operation), the amount of work done in going between specified initial
and final values in parameter space is in general path dependent.
From these results, we conclude that the
concepts of the thermodynamic potential and entropy when employed in a straightforward generalization
of the equilibrium formalism are of limited use in characterizing
the behavior of the system studied here. Apparently, a meaningful application of these concepts to the investigation of non-equilibrium
steady states will require the development of a radically novel approach.

\subsubsection{Thermodynamic quantities}

Here we introduce generalized thermodynamic quantities defined in such a manner to obtain a formalism analogous
to that of equilibrium thermodynamics \cite{fn5}.

Let us begin by reconsidering (\ref{changeofenergy}). Integrating this over the entire process under consideration,
we obtain the following:
\begin{equation}
\Delta U = W_F - \int_{t_{\rm{i}}}^{t_{\rm{f}}} dt \int dx F j^{\rm{non}} \; . \label{changeinenergy}
\end{equation}
Now, let us consider
the limit of a quasi-static operation. Then we have $\int_{t_{\rm{i}}}^{t_{\rm{f}}} dt w_F(t) \rightarrow
\int dx \int_{\alpha_{\rm{i}}}^{\alpha_{\rm{f}}} d\alpha  \frac{\partial V}{\partial \alpha} \rho_{\rm{ss}}(x ; \alpha)$.
This quantity, which we call $W_F^{\rm{ss}}$, is obviously independent of $\mu$ and is reversible. In other words, it depends only
on $\alpha_{\rm{i}}$ and $\alpha_{\rm{f}}$. For this reason, below we often refer to this quantity as the ``reversible work."
Then, because $\Delta U$ also depends only on $\alpha_{\rm{i}}$ and $\alpha_{\rm{f}}$, the
quantity $\lim_{\mu \rightarrow 0} \int_{t_{\rm{i}}}^{t_{\rm{f}}} dt \int dx F j^{\rm{non}}$ is a state function,
and it is natural to define the thermodynamic potential $\Phi(\alpha)$ through the relation
\begin{equation}
\Delta \Phi = \Delta U + \lim_{\mu \rightarrow 0} \int_{t_{\rm{i}}}^{t_{\rm{f}}} dt \int dx F j^{\rm{non}} \; . \label{potential1}
\end{equation}
In fact, if we wish for our thermodynamic potential to play the same role as that in the equilibrium theory -- i.e., the quantity whose change
represents the reversible work done on the system through the work coordinates -- then this is the unique choice.
It should also be noted here that for fixed $\alpha$, it is indeed the steady state solution that minimizes $\Phi$, and hence this
quantity truly is analogous to the equilibrium free energy.
This can be demonstrated by considering a variation of $\Phi$ and showing that the $\rho$ for which this variation
vanishes is also that for which the relation $j = \overline{j}$ holds. This calculation is made somewhat non-trivial
by the fact that $\Phi$ contains $j^{\rm{non}}$. For this reason,
when carrying out the variation, it is necessary to include the virtual variation of the current corresponding to a virtual
variation of the probability distribution.

Now, as shown in Appendix \ref{appent}, in the case of a quasi-static operation, the change in the entropy
of the system (defined as $S \equiv - \int dx \rho(x) \ln \rho(x)$) is given by
\begin{equation}
\Delta S = T^{-1} \lim_{\mu \rightarrow 0} \int_{t_{\rm{i}}}^{t_{\rm{f}}} dt \int dx 
\left[ \gamma \frac{j^{\rm{non}} \overline{ j }}{ \rho}  - F j^{\rm{non}} \right] \; . \label{changeinentropy}
\end{equation}
Then, combining this with (\ref{potential1}), we obtain the following:
\begin{equation}
\Delta \Phi = \Delta U - T \Delta S + \gamma \lim_{\mu \rightarrow 0} 
\int_{t_{\rm{i}}}^{t_{\rm{f}}} dt \int dx \frac{j^{\rm{non}} \overline{ j }}{ \rho}     \; . \label{potential2}
\end{equation}
Next, note that the quantity $\gamma \overline{ j } /\rho$ takes the form of a frictional force,
and thus $\gamma \int_{t_{\rm{i}}}^{t_{\rm{f}}}dt \int dx \frac{j^{\rm{non}} \overline{ j }}{ \rho}$ can be regarded as a type of frictional work.
Then, because $\Delta S$ itself is obviously reversible, the quantity $\gamma \lim_{\mu \rightarrow 0} 
\int_{t_{\rm{i}}}^{t_{\rm{f}}} dt \int dx \frac{j^{\rm{non}} \overline{ j }}{ \rho}$ represents a reversible frictional
work that characterizes a quasi-static transition between two non-equilibrium steady states.
Considering the form of (\ref{potential2}), it is natural to define the {\it generalized entropy}, $\Sigma$,
through the relation $\Delta \Sigma = \Delta S - \gamma T^{-1} \lim_{\mu \rightarrow 0} \int_{t_{\rm{i}}}^{t_{\rm{f}}}
dt \int dx \frac{j^{\rm{non}} \overline{ j }}{ \rho}$. For a system described by (\ref{equationofmotion}), in the case $f \ne 0$,
the quantity $\gamma T^{-1} \lim_{\mu \rightarrow 0} \int_{t_{\rm{i}}}^{t_{\rm{f}}}
dt \int dx \frac{j^{\rm{non}} \overline{ j }}{ \rho}$ is generally non-zero, and hence the generalized entropy
appearing in the thermodynamic potential differs from $S$. In the system considered here,
the difference between $S$ and $\Sigma$ appears at ${\cal O}(b^4)$, vanishing at lowest order [i.e.~${\cal O}(b^2)$].

Then, representing the value of $Q^{\rm{non}}$ in the quasi-static
limit by $Q^{\rm{non}}_{\rm{ss}}$, and comparing (\ref{potential1}) and (\ref{potential2}),
we obtain the generalized Clausius relation
\begin{equation}
\Delta\Sigma = Q^{\rm{non}}_{\rm{ss}} /T \; . \label{entropyandheat}
\end{equation}
Below, we often refer to $Q^{\rm{non}}_{\rm{ss}}$ as the ``reversible heat."

It is important to note the difference between $Q^{\rm{non}}$ appearing here and the so-called extra heat defined in \cite{oono2}.
In that work, it is conjectured that for an operation through which a system experiences a transition
from one non-equilibrium steady state to another, the change in the entropy is greater than or
equal $Q^{\rm{ex}}/T$, where $Q^{\rm{ex}}$ is the extra heat, with equality in the case of a
quasi-static operation. (For the definition of the extra heat, see Appendix \ref{appqex}.) Here it is noted
that for the system considered here, $Q^{\rm{ex}}$ and $Q^{\rm{non}}$ are distinct quantities both
for quasi-static and non-quasi-static operations. Further, for
this system, the conjecture made in \cite{oono2} does not hold. (For details, see Appendix \ref{appqex}.)
However, it should be noted that the manner of thinking presented in \cite{oono2}, in which the heat is divided into different types, is quite useful in the above analysis.

\subsubsection{Case of finite $\mu$}
\label{finitemu}

We have found that in the case of a quasi-static operation, the system considered here can be
described using a formalism similar to that of an equilibrium system, with $\Sigma$ playing the role of the entropy,
and $Q^{\rm{non}}$ playing the role of the heat. We now consider the case of finite $\mu$.

Using the ${\cal O}(\mu^1, b^1)$ form of $\rho(x,t)$ derived above, the change in energy of the system as a result
of the operation is obtained as
\begin{equation}
\Delta U = - \frac{k^2 T}{2\theta}[\alpha_{\rm{f}}^2 - \alpha_{\rm{i}}^2]c_0 L \; . \label{changeinenergy2}
\end{equation}
Then, from the relation $W_F = \int_{t_{\rm{i}}}^{t_{\rm{f}}} dt \rho \frac{dV}{dt}$, we derive the following:
\begin{equation}
W_F = \left[ - \frac{k^2 T}{4\theta}[\alpha_{\rm{f}}^2 - \alpha_{\rm{i}}^2]
 +\gamma \frac{k^2 T^2 -f^2}{2\theta^2} \int_{t_{\rm{i}}}^{t_{\rm{f}}} dt \dot{\alpha}^2 \right] c_0 L. \label{totalwork}
\end{equation}
Let us assume here that $t_{\rm{f}} - t_{\rm{i}}$ is ${\cal O}(\mu^{-1})$, so that the second term above is generically
${\cal O}(\mu)$.
Next, combining (\ref{changeinenergy2}) and (\ref{totalwork}) and employing the definition of $Q^{\rm{non}}$, we immediately obtain
\begin{equation}
Q^{\rm{non}} = \left[ - \frac{k^2 T}{4\theta}[\alpha_{\rm{f}}^2 - \alpha_{\rm{i}}^2]
 - \gamma \frac{k^2 T^2 -f^2}{2\theta^2} \int_{t_{\rm{i}}}^{t_{\rm{f}}} dt \dot{\alpha}^2 \right] c_0 L. \label{totalheat}
\end{equation}
Then, noting that the first term here (i.e., that independent of $\mu$) is $Q^{\rm{non}}_{\rm{ss}}$, and using (\ref{entropyandheat}),
we arrive at the following relation for $\Delta \Sigma$, valid to ${\cal O}(\mu^1, b^2)$:
\begin{equation}
\Delta \Sigma = \frac{ Q^{\rm{non}}}{T} + \gamma \frac{k^2 T^2 -f^2}{2\theta^2} c_0 L \int_{t_{\rm{i}}}^{t_{\rm{f}}} dt \dot{\alpha}^2 \; . \label{totalchangeinentropy}
\end{equation}
It is thus found that, to this order in $\mu$ and $b$, if $f > k T$, we have $\Delta \Sigma \le Q^{\rm{non}}/T$.
Then, although we have not explicitly computed the ${\cal O}(\mu)$ term in
$\Delta\Sigma$ to higher order in $b$, through a simple calculation we have found that
the change in sign of this term does {\it not} generally take place at  $f = k T$. It is thus concluded that the sign
of this ${\cal O}(\mu)$ term depends in some non-trivial manner on $T$, $f$ and $\alpha$. Hence, it is seen that the relation between
$\Delta \Sigma$ and $Q^{\rm{non}}$ for finite $\mu$ is much more complicated than that between the change of the
entropy and the transfer of heat in the equilibrium case, as expressed by the second law. The implication here is that the amount of work that
can be extracted from the system through the work coordinates in a transition between two steady states is not always maximized by a quasi-static operation.


Before moving on, we note that, as seen from the argument given in Appendix \ref{apptrans} (and the fact that
$\dot{\alpha}$ is ${\cal O}(\mu)$), the contribution to $W_F$ from
the transient behavior near $t = t_{\rm{i}}$ is ${\cal O}(\mu^2)$. Thus, (\ref{totalwork}) and (\ref{totalheat})
are indeed correct to ${\cal O}(\mu)$.

\subsubsection{Case of a two-component potential}

To this point, for simplicity, we have considered a potential $V$ that contains a single adjustable macroscopic
parameter. Here, however, we briefly examine the case in which there are two such parameters. We do this to
demonstrate that although quasi-static operations carried out on the type of systems studied here are reversible,
they are not generally path-independent.

Here, we consider (\ref{equationofmotion}) with $V(x;\alpha_1,\alpha_2) = \alpha_1 \cos(kx) + \alpha_2 \cos(2kx)$.
Then, we study operations in which both $\alpha_1$ and $\alpha_2$ are varied. We assume that
we can write $\alpha_j(t) = b a_j(t)$ for $j =1,2$, where $b$ is a small quantity and $a_j(t)$ is ${\cal O}(1)$ throughout these operations.
In this case, it is easily found that, to order $b^2$, the steady state solution
is given by
\begin{eqnarray}
\rho_{\rm{ss}}(x; \alpha_1, \alpha_2) = \left\{ 1 + \left[ - \frac{k^2 T}{\theta}\alpha_1 
+ \frac{4k^2}{\tilde{\theta}} \alpha_1 \alpha_2 \right] \cos(kx) \nonumber \right. \\
+ \left[ -\frac{4k^2T}{\tilde{\theta}} \alpha_2 + \frac{k^2(2k^2T^2 + f^2)}{\theta \tilde{\theta}} \alpha_1^2 \right] \cos(2kx) \nonumber \\
\left[ - \frac{k f}{\theta}\alpha_1 - \frac{6k^3fT}{\theta \tilde{\theta}} \alpha_1 \alpha_2 \right] \sin(kx) \nonumber \\
\left. + \left[ -\frac{2kf}{\tilde{\theta}} \alpha_2 + \frac{3k^3fT}{\theta \tilde{\theta}} \alpha_1^2 \right] \sin(2kx) \right\} c_0
\; , \label{twocomponent}
\end{eqnarray}
where $\tilde{\theta} = 4k^2T^2 + f^2$.
Then, using this form, we compute the work done by the external agent in two processes, one in which first $\alpha_1$
is changed from its initial value, $\alpha_1^{\rm{i}}$, to its final value, $\alpha_1^{\rm{f}}$, and then
$\alpha_2$ is changed from $\alpha_2^{\rm{i}}$ to $\alpha_2^{\rm{f}}$, and one in which this order is reversed.
In the first case, the total work, $W_{1,2}$, is given by
\begin{equation}
W_{1,2} = \int dx \int_{\alpha_1^{\rm{i}}}^{\alpha_1^{\rm{f}}} d \alpha_1
\rho_{\rm{ss}}(x; \alpha_1, \alpha_2^{\rm{i}}) \frac{\partial V(x; \alpha_1, \alpha_2^{\rm{i}})}{\partial \alpha_1} +
\int dx \int_{\alpha_2^{\rm{i}}}^{\alpha_2^{\rm{f}}} d \alpha_2
\rho_{\rm{ss}}(x; \alpha_1^{\rm{f}}, \alpha_2) \frac{\partial V(x; \alpha_1^{\rm{f}}, \alpha_2)}{\partial \alpha_2} \; , \label{w12}
\end{equation}
while in the second case, the total work, $W_{2,1}$,  is given by
\begin{equation}
W_{2,1} = \int dx \int_{\alpha_2^{\rm{i}}}^{\alpha_2^{\rm{f}}} d \alpha_2
\rho_{\rm{ss}}(x; \alpha_1^{\rm{i}}, \alpha_2) \frac{\partial V(x; \alpha_1^{\rm{i}}, \alpha_2)}{\partial \alpha_2} +
\int dx \int_{\alpha_1^{\rm{i}}}^{\alpha_1^{\rm{f}}} d \alpha_1
\rho_{\rm{ss}}(x; \alpha_1, \alpha_2^{\rm{f}}) \frac{\partial V(x; \alpha_1, \alpha_2^{\rm{f}})}{\partial \alpha_1} \; . \label{w21}
\end{equation}
Simple calculations yield
\begin{equation}
W_{1,2} - W_{2,1} =  -\frac{3k^2 f^2}{\theta \tilde{\theta}}
\left( {\alpha_1^{\rm{f}}}^2 - {\alpha_1^{\rm{i}}}^2 \right) \left( \alpha_2^{\rm{f}} - \alpha_2^i \right) L\;  .
\label{difference}
\end{equation}
In the equilibrium case, this difference vanishes, but
for non-zero $f$, it does not. It is thus seen that for the system considered here,
the work done in going from $(\alpha_1^{\rm{i}}, \alpha_2^{\rm{i}})$
to $(\alpha_1^{\rm{f}}, \alpha_2^{\rm{f}})$, even for quasi-static operations, depends on the path taken between them. Thus, although for a given path in
$\vec{\alpha}$-space, a quasi-static operation is reversible, and hence we can define a thermodynamic potential
and thus a generalized entropy for that path, this potential and this entropy are path dependent. Therefore, they
are not state functions.


\subsection{Conclusion regarding generalization of the equilibrium formalism}

The importance of the entropy in the study of equilibrium systems
is due to the simplicity of its properties, as described by the second law. This simplicity allows us to define a
thermodynamic potential that is useful because, first, it is a state function, and hence the work done by the system in all quasi-static operations connecting
two equilibrium states is uniquely
given by the difference between the values of the potential in these states, and, second, the work done by the
system in a non-quasi-static operation connecting the same two states is always less than this value.
By contrast, for the type of system considered in this paper, although we can define
a thermodynamic potential analogous to that for equilibrium systems, such a potential possesses neither of
these properties.

The results of this paper lead us to conjecture that the type of phenomena described by the equilibrium theory,
i.e., the interplay between thermal and mechanical energy in operations connecting steady states, cannot  be described
systematically in the non-equilibrium case. This would seem to imply that the entropy and the thermodynamic
potential will not play central roles in a general phenomenological theory of non-equilibrium steady states.
We are thus led to seek a theoretical framework for such states based on some other quantities.

\section{Conclusion}
\label{conc}

The broad goal of thermodynamics is to construct descriptions of macroscopic
behavior in terms of macroscopic observables. The problem encountered in this pursuit is
that it is generally difficult
within a purely macroscopic description to account for the effect that microscopic degrees
of freedom have on the macroscopic observables. In the case
of a system in equilibrium, very large numbers of microscopic degrees of freedom act collectively in such a way that
they can be treated as a small number of macroscopic degrees of freedom. For this reason, when a system undergoes an
infinitesimally slow operation, during which it is always infinitesimally close to an equilibrium state, a simple,
quasi-mechanical framework based on a thermodynamic potential expressed entirely in terms of macroscopic state variables
is sufficient to describe the total amount of thermal energy converted into mechanical energy in the transition between some initial
and final states.
Then, considering the set of all operations connecting these states, the amount of mechanical
energy thus obtained is maximal in this case of an infinitesimally slow operation. These simple properties form the basis of the
thermodynamic theory of equilibrium states. 
In this paper, we have shown how the attempt to formulate an analogous theory of non-equilibrium steady states
-- namely, a quasi-mechanical model describing the interplay between thermal and mechanical forms of energy during
externally applied operations --
meets with difficulty in the case of a particular model. 
As discussed in the previous section, the difficulty encountered here results from the complicated nature of the work done by the system in
operations (even quasi-static operations) connecting non-equilibrium steady states.

As stated above, the fundamental problem in constructing any type of thermodynamic formalism is to determine how to account for
the effect of microscopic degrees of freedom on macroscopic behavior through expressions involving macroscopic observables.
Certainly, for any given system, there are limitations on the types of behavior that can be described simply in such a
phenomenological manner. Thus, in the attempt to construct a phenomenological theory for a given class of systems, it is of fundamental importance
to ascertain the types of behavior for which simple phenomenological descriptions are feasible.
The results of the present work suggest that the conversion of heat into mechanical energy, which is the focus of the thermodynamic theory of equilibrium states, may be
of a type for which a general phenomenological description is unfeasible in the case of non-equilibrium steady states.
This leads us to conclude that in order to construct a phenomenological theory of non-equilibrium steady states,
we will need radically new concepts involving novel quantities. It is our belief that the search for non-equilibrium theories should now be focused on finding such
concepts and quantities, rather than attempting to extrapolate from the equilibrium theory.

\begin{acknowledgments}
I am grateful to T.~Ohta and S.-i.~Sasa for helpful discussions. This work was supported by the Grant-in-Aid for the Global COE Program
``The Next Generation of Physics, Spun from Universality and Emergence" from the Ministry of Education, Culture, Sports, Science and Technology (MEXT) of Japan.
\end{acknowledgments}

\appendix

\section{Perturbative solution to equation of motion}
\label{apppert}

In this section we derive the ${\cal O}(\mu^1, b^1)$ solution to (\ref{equationofmotion}) using a renormalization
group (RG) method.
The formulation of RG employed here is presented in \cite{merg}. This formulation is designed
to treat differential equations in which there appear quantities that exhibit slow variation with respect to
the independent variables. In the simplest situation, such an equation describes
a system possessing some kind of slowly varying temporal and/or spatial inhomogeneity, and this is represented
by slowly varying explicit functions of time and/or space contained in the equation.
With this method, the system is first treated locally
by expanding the slowly varying quantities in the equation about some fixed values of the independent 
variables and then deriving a locally valid solution to the resulting local equation. Then, the RG
procedure is used to obtain a globally valid solution by appropriately connecting all such local solutions.

In the present case, the slowly varying quantity is $\alpha(\mu t) = b a(\mu t)$. Thus, as the first step in solving
(\ref{equationofmotion}),
we expand $\alpha$ about some arbitrary value of time, $t_0$. Because we seek the ${\cal O}(\mu^1)$
solution to the original equation of motion, we retain only terms in the local equation up to ${\cal O}(\mu^1)$. Thus,
we consider the following:
\begin{eqnarray}
\gamma \dot{\tilde{\rho}}^{(1)}(x,t) &=& \left\{ T \frac{\partial^2 }{\partial x^2} - \left[ {f} +
b \left({a}(\mu t_0)  + \mu (t - t_0) a^\prime (\mu t_0) \right) k \sin(kx) \right]\frac{\partial}{\partial x} \right. \nonumber \\
&& \left. - b  \left( {a} (\mu t_0) + \mu (t -t_0)a^\prime \right) k^2 \cos (kx) \right\} \tilde{\rho}^{(1)}(x,t) \; .
\label{rg1}
\end{eqnarray}
Here, we have $a^\prime \equiv d a/d (\mu t)$.
Also note that throughout this section, the tilde indicates that
the quantity in question is defined only locally, and a superscript $(n)$
indicates that it is valid to ${\cal O}(\mu^n)$. 

Next, we solve the local equation (\ref{rg1}) order by order in $\mu$, explicitly constructing an ${\cal O}(\mu^1)$ 
solution of the form $\tilde{\rho}^{(1)}(x,t) = \tilde{\rho}_0(x,t) + \mu \tilde{\rho}_1(x,t)$.

First, to obtain $\tilde{\rho}_0(x,t)$, we keep only the ${\cal O}(\mu^0)$ terms in (\ref{rg1}). This yields the equation
\begin{equation}
\gamma \dot{\tilde{\rho}}_0(x,t) = \left\{ T \frac{\partial^2 }{\partial x^2}
- \left[ {f} + b {a}(\mu t_0) k \sin(kx) \right] \frac{\partial}{\partial x}
- b  {a} (\mu t_0) k^2 \cos (kx) \right\} \tilde{\rho}_0(x,t) \; .
\label{rg2}
\end{equation}
Now, because we seek solutions to (\ref{equationofmotion}) that are near steady state solutions \cite{fn6},
we are interested in the solution to (\ref{rg2}) satisfying $\dot{\tilde{\rho}}_0(x,t) = 0$.
Using this, we obtain the following ${\cal O}(b^1)$ solution to this equation:
\begin{equation}
\tilde{\rho}_0(x,t) = c_0 \left[ 1 + b {a}(\mu t_0) \left( \sigma^0_+ \cos(kx) + \sigma^0_- \sin (kx) \right) \right] \; .
\label{rg3}
\end{equation}
The quantities $\sigma^0_+$ and $\sigma^0_-$ appearing here are given in Section 3. The function in (\ref{rg3})
represents a solution valid to ${\cal O}(\mu^0, b^1)$ in the neighborhood of $t = t_0$.
Then, to obtain the global ${\cal O}(\mu^0, b^1)$ solution, we merely set $t_0 = t$ in (\ref{rg3}). This gives the
lowest-order solution presented in Section 3,
\begin{equation}
\rho^{(0)}(x,t) = c_1 \left[ 1 + b {a}(\mu t) \left( \sigma^0_+ \cos(kx) + \sigma^0_- \sin (kx) \right) \right] \; .
\label{rg4}
\end{equation}
The ${\cal O}(\mu^0)$ RG equation, obtained by retaining the ${\cal O}(\mu^0)$ terms
in the equation $\frac{\partial \tilde{\rho}_0(x,t)}{ \partial t_0}|_{t_0 = t} = 0$,
is simply $0= 0$.
As stated in Section 3, if we replace $\alpha(\mu t)$ by some fixed value $\alpha$, the solution given in (\ref{rg4})
represents the steady state solution to the original equation of motion at this value of $\alpha$.

Next, we derive $\tilde{\rho}_1(x,t)$. This is done by first extracting the ${\cal O}(\mu^1)$ terms in (\ref{rg1})
and thereby constructing the following:
\begin{eqnarray}
\gamma \dot{\tilde{\rho}}_1(x,t) &=& \left\{ T \frac{\partial^2 }{\partial x^2} - \left[ {f} +
b {a}(\mu t_0) k \sin(kx)\right] \frac{\partial}{\partial x}
- b  {a} (\mu t_0) k^2 \cos (kx) \right\} \tilde{\rho}_1(x,t) \nonumber \\
&-& \left\{ b (t - t_0) a^\prime(\mu t_0)  \left[k \sin(kx) \frac{\partial }{\partial x}
+ k^2 \cos(kx) \right] \right\} \tilde{\rho}_0(x,t) \; .
\label{rg5}
\end{eqnarray}
The ${\cal O}(b^1)$ solution to this equation is given by
\begin{equation}
\tilde{\rho}_1(x,t) =
c_0 \gamma a^\prime(\mu t_0)
\left\{
	b \left[
				 \sigma^1_+ \cos(kx) + \sigma^1_- \sin (kx) 
		\right]
+ (t - t_0) 
		\left[ - b  \sigma^0_+ \cos(kx) + \sigma^0_- \sin(kx) 
			 \right] \right\}\; .
\label{rg6}
\end{equation}

We have thus derived the local solution valid to ${\cal O}(\mu^1, b^1)$, $\tilde{\rho}^{(1)}(x,t) = \tilde{\rho}_0(x,t) + \mu \tilde{\rho}_1(x,t)$.
Then, setting $t_0 = t$ in this form, we obtain the global solution valid to the same order:
\begin{eqnarray}
\rho^{(1)}(x,t) = c_0 \left\{ 1 + b a(\mu t) \left[ \sigma^0_+ \cos (kx) + \sigma^0_- \sin (kx) \right] \right. \nonumber \\ \left.
+ \mu \gamma b a^\prime(\mu t) \left[ \sigma^1_+ \cos(kx) + \sigma^1_- \sin (kx) \right] \right\} \; .
\label{rg7}
\end{eqnarray}
At this order, the RG equation is given by $\frac{\partial \tilde{\rho}^{(1)}(x,t)}{ \partial t_0}|_{t_0 = t} = 0$.
Solving this, we obtain $\dot{\alpha}(t) = \dot{\alpha}(t)$, and, as in the ${\cal O}(\mu^0)$ case, it is trivial \cite{fn7}.

The following form of $\rho^{(1)}(x,t)$ is more transparent:
\begin{equation}
\rho^{(1)}(x,t) = c_0 \left\{ 1 + \left[  \sigma^0_+ \alpha + \gamma \sigma^1_+ \dot{\alpha} \right]\cos(kx) +
\left[ \sigma^0_- \alpha + \gamma \sigma^1_- \dot{\alpha} \right] \sin (kx) \right\} \; .
\label{rg8}
\end{equation}
The form of this solution represents the ${\cal O}(\mu^1, b^1)$ approximation
to the RG fixed point (or center manifold) of (\ref{equationofmotion}). The slow motion contained within this form,
coming from the time dependence of $\alpha$ and $\dot{\alpha}$, represents the dynamics on this fixed point (or manifold).

\section{Estimation of the contribution of transient regimes}
\label{apptrans}

As noted in Section \ref{pert}, the solution derived there describes the system in the `scaling' regime,
after transient behavior has disappeared. However, in several places in this paper, we calculate quantities
by integrating over the time interval $[t_{\rm{i}}, t_{\rm{f}}]$, assuming this solution to be valid for
all such times. The fact that we ignore the transient behavior in these calculations introduces an error.
Here we present an argument from which the size of this error can be estimated.

For the type of experiments considered here, there are two intervals during which transient behavior will appear,
one for $t$ slightly greater than $t_{\rm{i}}$, as $\rho(x,t)$ relaxes to the form derived in Section \ref{pert},
and one for $t$ slightly greater than $t_{\rm{f}}$, as $\rho(x,t)$ relaxes to the final steady state.
First we consider the behavior for $t$ near $t_{\rm{i}}$.

At $t= t_{\rm{i}}$, the difference, $\delta \rho(x,t)$, between the actual $\rho(x,t)$
and that given by (\ref{order1form}) is ${\cal O} (\mu)$. (For the purpose of the present discussion,
let us refer to the latter as $\rho_{\rm{scal}}(x,t)$.) We seek to describe the time evolution of $\delta \rho(x,t)$.
This can be done by employing the computational technique demonstrated in Appendix \ref{apppert}. First, we
expand (\ref{equationofmotion}) locally in time about some $t_0$ near $t_{\rm{i}}$.
Then, to ${\cal O}(\mu)$, the (local) equation of motion for $\delta \rho(x,t)$ is
\begin{equation}
\gamma \frac{\partial {\delta \rho} (x,t)}{\partial t} = \left[ T \frac{\partial^2 }{\partial x^2}
- \left( f + F(x;\alpha(\mu t_0)) \right) \frac{\partial}{\partial x}
- \frac{\partial F(x;\alpha(\mu t_0))}{\partial x} \right] \delta \rho(x,t) \; . \label{trans1}
\end{equation}
Now, with the assumption that $\rho_{\rm{scal}}(x,t)$ represents a stable fixed point of the time evolution of (\ref{equationofmotion}),
one eigenfunction of this equation will have vanishing eigenvalue, and the rest will have negative eigenvalues.
That with eigenvalue 0 corresponds to $\rho_{\rm{scal}}(x,t)$ itself. Then, as (\ref{trans1})
contains only $O(1)$ quantities, all such negative eigenvalues will
be $O(1)$. Thus, when viewed locally in time, each mode (other than that corresponding to
$\rho_{\rm{scal}}(x,t)$)
will exhibit simple exponential decay with an ${\cal O}(1)$ decay constant. Let us consider an
arbitrary such mode and refer to its decay constant as $\lambda$.
Then, ``renormalizing" the local solution for this mode
(i.e., applying the RG equation and setting $t_0 = t$) to properly connect the local behavior corresponding to
each time $t_0$ (see \cite{merg}
for detailed discussion), we obtain solutions of the form $C \exp[-\lambda(\mu t)t]$ for each decaying mode.
(Thus, the decay constant $\lambda$ becomes a slowly varying function of time. This is due to
the fact that the local eigenvalue, $\lambda$, will contain $\alpha(\mu t_0)$, and hence depend on
the time $t_0$ about which the equation of motion is expanded.)
Thus, each mode will decay exponentially
with a slowly varying [but always $O(1)$] time ``constant." 
(This exponential decay represents the convergence of $\rho(x,t)$
to the center manifold, constituted by $\rho_{\rm{scal}}(x,t)$.
The slow variation in time of its decay constant is due to the
slow motion on the center manifold.) Thus, we have fast exponential decay of $\delta \rho(x,t)$
to zero, with the speed of this decay varying slowly in time. Therefore,
any integral $\int dx \int_{t_{\rm{i}}}^{t_{\rm{f}}} dt g(x,t) \delta \rho(x,t)$, where $g(x,t)$ is
an arbitrary ${\cal O}(1)$ function, will be ${\cal O}(\mu)$.

The situation regarding the transient after $t_{\rm{f}}$ is similar, but slightly simpler.
In that case, each mode (except that corresponding to the center manifold, which in this case
corresponds to $\rho_{\rm{ss}}(x;\alpha_{\rm{f}})$) will exhibit simple exponential
decay with an ${\cal O}(1)$ decay constant that is actually constant.

\section{Derivation of (\ref{changeinentropy})}
\label{appent}

Here we give the simple computations yielding (\ref{changeinentropy}).

Consider an arbitrary process (which, for example, may result from an externally applied
operation) taking place in the system described by (\ref{equationofmotion})
over some time interval $[t_{\rm{i}}, t_{\rm{f}}]$. For such a process,
we wish to find a relation between the change in the Shannon entropy,
$S \equiv - \int \rho(x,t) \log \rho(x,t) dx $,
and experimentally measurable quantities. In this process, the change in entropy is
given by
\begin{equation}
\Delta S =  - \int_{t_{\rm{i}}}^{t_{\rm{f}}} dt \int 
\left[ \dot{\rho}(x,t) \log \rho(x,t) + \dot{\rho}(x,t) \right] dx\; .
\end{equation}
Obviously, the spatial integral of the second term vanishes.
Thus, with the relation $\dot{\rho}(x,t) = - \nabla \cdot j(x,t)$,
this becomes
\begin{equation}
\Delta S =  \int_{t_{\rm{i}}}^{t_{\rm{f}}} dt \int \log [\rho(x,t)] \nabla \cdot j(x,t) dx \; .
\end{equation}
Then, using integration by parts, and the fact that we stipulate periodic boundary conditions, we obtain
\begin{equation}
\Delta S =  - \int_{t_{\rm{i}}}^{t_{\rm{f}}} dt \int   j(x,t) \cdot \nabla  \log[ \rho(x,t)] dx \; .
\label{ds3}
\end{equation}

Next, using the relation
\begin{equation}
j(x,t) = -  \gamma^{-1} T \nabla \rho(x,t) + \gamma^{-1} {\cal F}(x,t) \rho(x,t) 
\end{equation}
and writing $j(x,t)$ as $j(x,t) = \overline{j}(t) + j^{\rm{non}}(x,t)$,
we obtain the following:
\begin{equation}
\Delta S =  D^{-1} \int_{t_{\rm{i}}}^{t_{\rm{f}}} dt \int dx
j^{\rm{non}}(x,t) \left[ \frac{j^{\rm{non}}(x,t)}{\rho(x,t)} +
\frac{\overline{j}(t) }{\rho(x,t)}
- \gamma^{-1}F(x,t) \right] dx \; .
\label{ds5}
\end{equation}

Then, assuming that in the $\mu \rightarrow 0$ limit, $j^{\rm{non}}$ is ${\cal O}(\mu)$, we find
\begin{equation}
\Delta S =  T^{-1} \lim_{\mu \rightarrow 0} \int_{t_{\rm{i}}}^{t_{\rm{f}}} dt \int dx \left[
\gamma \frac{j^{\rm{non}}(x,t) \overline{j}(t)}{\rho(x,t)} - F(x,t)j^{\rm{non}}(x,t) \right] dx \; .
\label{ds6}
\end{equation}
This is the relation given in (\ref{changeinentropy}).
With $\Delta \Sigma$ as defined in the main text, (\ref{ds6}) is equivalent to $T\Delta \Sigma = Q^{\rm{non}}_{\rm{ss}} \equiv
- \lim_{\mu \rightarrow 0}  
\int_{t_{\rm{i}}}^{t_{\rm{f}}} dt \int dx F(x,t)j^{\rm{non}}(x,t)$.

\section{$Q^{\rm{non}}$ and $Q^{\rm{ex}}$}
\label{appqex}

In this appendix, we briefly compare the quantities $Q^{\rm{non}}$ and $Q^{\rm{ex}}$ and show why they differ
for the system studied here.

Following Oono and Paniconi, the housekeeping heat for an operation of the type considered in this
paper is defined as the total amount of heat that would be transferred
from the heat bath if at each time $t \in [t_{\rm{i}},t_{\rm{f}}]$, the system were in the steady state characterized by
$\alpha(t)$. More precisely, we have $Q^{\rm{HK}} \equiv \int_{t_{\rm{i}}}^{t_{\rm{f}}} dt q^{\rm{ss}}[\alpha(t)]$, where
$q^{\rm{ss}}[\alpha(t)]$ is the rate of heat production in the steady state corresponding to $\alpha = \alpha(t)$.
Then, the extra heat is defined as $Q^{\rm{ex}} \equiv Q - Q^{\rm{HK}}$.
Clearly, from these definitions, we have the relation $Q^{\rm{ex}} = - \int_{t_{\rm{i}}}^{t_{\rm{f}}} dt \int dx [F(x; \alpha(t)) + f]j^{\rm{ex}}(x,t)$,
where the extra current, $j^{\rm{ex}}(x,t)$, is defined as $j^{\rm{ex}}(x,t) \equiv j(x,t) - j^{\rm{ss}}(\alpha(t))$.
With a simple calculation, using the ${\cal O}(\mu^1, b^2)$ solution to (\ref{equationofmotion}) (which is not given
explicitly here) and the relation (\ref{current}),
we find, to ${\cal O}(\mu^1, b^2)$,
\begin{eqnarray}
&&j^{\rm{ex}}(x,t) =  \dot{\alpha}
\left\{
 \left[ (f \sigma^1_+ - k T \sigma^1_-) \cos(kx) + (kT \sigma^1_+ + f \sigma^1_-) \sin (kx) \right]
\right. \nonumber \\ 
&& + \left.
 \alpha \left[ (f \sigma^1_{2+} - 2kT \sigma^1_{2-}  - \frac{k \sigma^1_-}{2})\cos(2kx)
 + (2 kT \sigma^1_{2+} + f \sigma^1_{2-} + \frac{k \sigma^1_+}{2}) \sin(2kx) + \frac{k \sigma^1_-}{2} \right]
\right\}c_0 \; .
\label{extracurrent}
\end{eqnarray}
Here, the quantities $\sigma^1_{2+}$ and $\sigma^1_{2-}$ are the ${\cal O}(\mu^1, b^2)$ coefficients of the terms $\cos(2kx)$ and $\sin(2kx)$
in the solution $\rho(x,t)$. Because they are somewhat complicated, and the terms in which they appear in (\ref{extracurrent})
give no contribution to $Q^{\rm{ex}}$, we do not present them here.

Note that in the above expression for $j^{\rm{ex}}$, we have the spatially uniform component
$\frac{1}{2} \dot{\alpha} \alpha k \sigma^1_-c_0$. We refer to this as the ``uniform extra current," and write it
$\overline{j^{\rm{ex}}}(t)$.
Now, substituting the form in (\ref{extracurrent}) into the above equation for $Q^{\rm{ex}}$, we obtain the following ${\cal O}(\mu^0)$
expression for this quantity, which we refer to as $Q^{\rm{ex}}_{\rm{ss}}$:
\begin{equation}
Q^{\rm{ex}}_{\rm{ss}} =  - c_0\frac{T}{4} \left({\sigma^0_+}^2 + 3 {\sigma^0_-}^2\right)\left[\alpha_{\rm{f}}^2 - \alpha_{\rm{i}}^2\right] L\; .
\label{extra}
\end{equation}
This expression is valid to ${\cal O}(b^2)$.

Next, note that $j^{\rm{non}}(x,t) \equiv j(x,t) - \overline{j}(t)$ is given by $j^{\rm{non}}(x,t) = j^{\rm{ex}} - \overline{j^{\rm{ex}}}(t)$.
With this, we find, to ${\cal O}(b^2$),
\begin{equation}
Q^{\rm{non}}_{\rm{ss}} =  - c_0\frac{T}{4} \left({\sigma^0_+}^2 + {\sigma^0_-}^2\right)\left[\alpha_{\rm{f}}^2 - \alpha_{\rm{i}}^2\right] L\; .
\label{non}
\end{equation}
A straightforward computation shows that this is indeed equal to $T\Delta \Sigma$.
(Note that to this order in $b$, $\Delta S$ and $\Delta \Sigma$ are identical. Their difference appears at ${\cal O}(b^4)$.)

It is thus seen that $Q^{\rm{non}}$ and $Q^{\rm{ex}}$ differ as a result of the presence of the uniform extra current.
Also, it should be noted that both of these quantities are readily measured experimentally, and hence they can be directly compared.

Intuitively, it is reasonable that it would be $Q^{\rm{non}}$ and not $Q^{\rm{ex}}$ that is closely related to the change
in the entropy, as the description of the system (i.e.~$\rho$)
is obviously unchanged by any uniform current. In other words, it is the heat generated in the rearrangement of the probability
distribution necessary to change it from $\rho_{\rm{ss}}(x; \alpha_{\rm{i}})$ to $\rho_{\rm{ss}}(x;\alpha_{\rm{f}})$
that should be related to the change in entropy, and this heat is $Q^{\rm{non}}$.

It is important to note here that we have ignored the contributions of the transient behavior to $Q^{\rm non}$ and $Q^{\rm ex}$ for both $t$ slightly greater
than $t_{\rm{i}}$ and slightly greater than $t_{\rm{f}}$. However, from the argument given in Appendix \ref{apptrans},
it is seen that these contributions are no greater than ${\cal O}(\mu)$. Next, note
that $Q^{\rm{non}}_{\rm{ss}}$ can also be obtained from $\rho_{\rm{ss}}(x; \alpha(t))$ alone,
without deriving $j^{\rm{non}}$. This is seen from the relation $Q^{\rm{non}}_{\rm{ss}} = \Delta U - W_F^{\rm{ss}}$,
and noting that
$W_F^{\rm{ss}} = \int dx \int_{\alpha_{\rm{i}}}^{\alpha_{\rm{f}}}d \alpha \rho_{\rm{{ss}}}(x; \alpha) \frac{\partial V(x;\alpha)}{\partial \alpha}$,
as mentioned in the main text. Using this method of calculating $Q^{\rm{non}}_{\rm{ss}}$, it is seen (as discussed in \ref{pert})
that the contribution of the transient to $Q^{\rm{non}}$ is actually only ${\cal O}(\mu^2)$.

Finally, we point out that the result found here appears to be inconsistent with that obtained in \cite{hatano}. The
reason for this is that the definition of the extra heat there (see Eq.~(16) of that work)
is inconsistent with that studied here. It should be noted that the extra heat used in \cite{hatano} would be quite difficult to measure in all but
the simplest systems.

\section{Validity of the Langevin and Smoluchowski equations}
\label{applangevin}

Let us now consider the validity of the Smoluchowski equation, (\ref{equationofmotion}). Because this
equation can be derived quite generally from the Langevin equation (\ref{langevin}), it is sufficient
to consider the validity of the latter.

In this appendix, we discuss the validity of the Langevin equation taking the form
\begin{equation}
M \ddot{x}(t) = 
- \gamma \dot{x}(t) + {\cal F}(x(t);\alpha(t)) + \xi(t) \; ,
\label{langevin2}
\end{equation}
where, $M$ is the mass of the Brownian particle.
In the standard treatment, we have $\gamma = 6 \pi R \eta$ (where $\eta$ is the viscosity
and $R$ is the radius of the Brownian particle), and
the intensity of the stochastic force is given by $K = 2 \gamma T$, as determined by the fluctuation-dissipation relation.

The above equation is valid in application only to systems that possess two distinct types of motion, microscopic and macroscopic,
whose characteristic timescales are greatly separated.
For the present purposes, the relevant timescale of the microscopic motion, $\tau_{\rm micro}$, is the average time between collisions of the heat bath particles with
the Brownian particle. Then, there are two relevant timescales of the macroscopic motion, namely, the times required for the position
and the velocity of the Brownian particle to change by macroscopically measurable amounts, which, of course, depend on the spatial and temporal resolutions of the observation.
Let us refer to these as $\tau_{\rm macro \; 1}$ and $\tau_{\rm macro \; 2}$.
The main condition necessary for the dynamics of the Brownian particle to be describable by a Langevin equation of the form appearing in (\ref{langevin2}) is
that $\tau_{\rm macro \; 1}$ and $\tau_{\rm macro \; 2}$ be sufficiently longer than $\tau_{\rm micro}$ that the dynamics of the heat bath particles can be clearly separated into
fast, stochastic motion and slow, systematic motion, with the statistics of the fast motion being completely independent of the slow motion and the slow motion determined entirely by
the motion of the Brownian particle.
%
If this is the case, then there should be a fourth timescale, $\tau_{\rm local \; eq}$, satisfying $\tau_{\rm micro} \ll \tau_{\rm local \; eq} \ll \tau_{\rm macro \; 1}, \tau_{\rm macro \; 2}$,
on which the stochastic behavior of the heat bath particles realizes a stationary distribution, which at any time $t$
depends only on the position and velocity of the Brownian particle at this time, $x(t)$ and $\dot{x}(t)$, the global properties of the heat bath (e.g., the temperature,
the average particle concentration, and the mass of the heat bath particles) and the size and shape of the Brownian particle. If there exists no such time scale, then
the statistics of the force exerted by the heat bath on the particle at any time will depend in some non-trivial manner on the history of the particle's motion, its interaction with the
heat bath will depend non-linearly on its velocity, the stochastic force it experiences will be non-Gaussian, and in general, formulating a meaningful description of its motion will be hopeless.
Now, with the simplest considerations, we are able to derive the following set of conditions which must be met in
order for such a timescale to exist:
$\dot{x}$ must be much smaller than both $(R/l)^{1/2} \overline{c}$
and $c_{\rm{sound}}$ (where $l$ is the mean free path of the heat bath particles, $\overline{c}$ is the rms velocity of the heat bath particles, and $c_{\rm sound}$ is the sound
velocity in the heat bath), and the timescale over which $\dot{x}$ varies must be
much longer than both $R / \overline{c}$ and $l / c_{\rm{sound}}$.

Next, we consider the conditions on the timescale of the description, $\tilde{\tau}$, that must be met in order for the Langevin equation to be valid. First, we note that there are several other
important timescales that characterize the system in addition to those cited above. First, there is the
timescale on which the Brownian particle experiences a set of collisions that are representative of the stationary distribution of the heat bath particles.
It should be the case that this timescale and $\tau_{\rm local \; eq}$ are of the same order (although, certainly, the former must be longer than the latter), and for our purposes,
they can be regarded as the same.
Next, there is the correlation time of the Brownian particle's velocity, $\tau_{\rm corr}$. Then, there is the timescale
over which the force ${\cal F}(x(t); \alpha(t))$ changes, $\tau_{\cal F}$.
The first condition on $\tilde{\tau}$ is that it must satisfy $\tilde{\tau} \gg \tau_{\rm local \; eq}$.
If this is not the case, then there will be no well-defined friction constant nor intensity of the stochastic force \cite{fn8}.
Second, it must be the case that $\tilde{\tau} \ll \tau_{\rm macro \; 1}$,
in order for the trajectory of the particle to be continuous. That we can choose a $\tilde{\tau}$ that satisfies these two conditions follows from the main condition given above.
Third, it must be the case that $\tilde{\tau} \ll  \tau_{\cal F}$, in order for the force in the equation to be well-defined. Of course, this third condition follows from the second as long
as ${\cal F}$ can be regarded as constant on lengthscales smaller than the resolution of the observation and $\alpha$ can be regarded as constant on timescales shorter than
$\tilde{\tau}$.

Finally, assuming that the above conditions are all satisfied, and the Langevin equation provides a proper description of the system's dynamics, then in order for these dynamics to be
overdamped, it must be the case that $\tau_{\rm corr} \ll  \tau_{\cal F}$. (Note that here, we can regard $\tau_{\rm corr}$ as $M/\gamma$.) This
implies the relation $\dot{x}_{\rm rms} \gg |\frac{m}{\gamma^2} \left[ \dot{\alpha} \frac{\partial {\cal F}(x)}{\partial \alpha} + \dot{x} \cdot \nabla {\cal F}(x) \right]|$, where $\dot{x}_{\rm rms}$
is the root-mean-square velocity of the Brownian particle, as measured over the timescale $\tau_{\rm corr}$. This condition is sufficient for an equation of the form (\ref{langevin}) to provide
a valid description of the dynamics. However, for this form with a delta-function correlated noise to be valid, it is also necessary that we have $\tilde{\tau} \gg \tau_{\rm corr}$. 
Then, in order for the position of the particle to be well-defined at all times with such a description, it is necessary that  we have $|\dot{x}(t)| \tau_{\rm corr}  \ll \Delta x$, where $\Delta x$ is 
spatial resolution of the experimental measurements.


\begin{thebibliography}{99}
\bibitem{fn1}
These works are discussed briefly in \cite{me-current}.
\bibitem{fn1.1}
More accurately, the `system' here is represented by the projection of the phase point in the 2-dimensional phase space onto the 1-dimensional configuration
space. Hence, the momentum fluctuations of the particle are regarded as part of the heat bath.
\bibitem{fn2}
In fact, this Smoluchowski equation is more generally valid than the overdamped Langevin equation given in (\ref{langevin}).
\bibitem{fn2.1}
More precisely, in general, the current $j$ considered here is defined as $j(x,t) \equiv \int dp \frac{p}{M} \rho(p,x,t)$, where $p$ is the momentum of the
particle, $M$ is its mass, $\rho(p,x,t)$ is the probability distribution in the phase space, and the integral is over all values of $p$.
\bibitem{fn3}
Note
that this depends on the assumption implicit in the Langevin equation that the heat bath is infinitely
large and hence that its temperature is unchanged by the collisions of the heat bath particles with the Brownian
particles. For a heat bath of finite size, such a strong claim cannot be made.
Further discussion of the conditions necessary for the validity of the Langevin equation is given in Appendix \ref{applangevin}.
\bibitem{fn4}
The
conditions on the existence of such a solution for (\ref{equationofmotion}) are quite weak. The physical conditions necessary
for this equation to provide a valid description of the system under consideration are stronger. Essentially, it is required that
$\alpha$ be changed sufficiently slowly that the dynamics of the system remain overdamped. This and related points are
discussed in Appendix \ref{applangevin}.
\bibitem{fn5}
We note here that all of the above analysis is very easily generalized to the case of $N$ interacting Brownian particles in a three-dimensional system.
In that case, we consider the system in the $3N$-dimensional configuration space. Then,
all of the above expressions hold if we replace $\overline{j}$ by the current in the divergenceless subspace (defined as the space consisting
of all vector fields in $\Re^{3N}$ whose divergence vanishes everywhere), $j^{\rm non}$ by that orthogonal to this subspace, and $q^{\rm non}$
by the rate of work done by the heat bath on the latter component of the current.
\bibitem{fn6}
 Hence, we seek
the behavior of (\ref{equationofmotion}) that is structurally stable with respect to the perturbation yielding
the time dependence of $\alpha$.
\bibitem{fn7}
The
reason that the RG equation is trivial here is that the slow motion exhibited by the solution above the bifurcation at $\mu = 0$
is due entirely to the resulting explicit time dependence of the equation of motion.
\bibitem{fn8}
Note that this condition is unnecessary in the case of the
Smoluchowski equation. Thus, the description given by this equation applies to shorter timescales than that given by the Langevin equation.
\end{thebibliography}
\end{document}